# Superconductivity and Aluminum Ordering in $Mg_{1-x}Al_xB_2$

J.Q. Li[a], L. Li[a,b], F. M. Liu[a,c], C. Dong[a], J. Y. Xiang[a], and Z.X. Zhao[a]

[a] National Laboratory for Superconductivity, Institute of Physics, Chinese Academy of Sciences, Beijing 100080, P. R. China

[b] Department of Physics, Ningxia University, Yinchuan 750021, P.R. China.

[c] Center for Material Physics and Chemistry, Beijing University of Aeronautics & Astronautics, Beijing 100083, P. R. China

Superconductivity and structural properties of $Mg_{1-x}Al_xB_2$ materials have been systematically investigated. Evident modifications in both superconductivity and microstructure are identified to originate from the Al ordering along the c-axis direction. The resultant superstructure phase has an optimal composition of $MgAlB_4$ with the superconducting transition at around 12K. Brief diagrams illustrating the superconductivity and structural features of $Mg_{1-x}Al_xB_2$ materials along with the increase of Al concentration are presented.

----------------
Dr. Li Jianqi
National Laboratory for Superconductivity,
Institute of Physics, Chinese Academy of Sciences
Beijing 100080, P.R. China.
Email: ljq@ssc.iphy.ac.cn

Magnesium diboride ($MgB_2$), as a phonon-mediated high-Tc superconductor, has attracted considerable interest from both theoretical and experiential points of view [1-5]. The superconducting transitions in the $MgB_2$ materials synthesized under a variety of conditions seem to be at around the limit of $T_c$ as suggested theoretically several decades ago for BCS. Recent theoretical calculations show that the strong coupling in the $MgB_2$ superconductor originates fundamentally from the in-plane optical phonon related to the vibrations of B atoms [6, 7]. Doping Al on the Mg site can introduce electrons into the bands and, moreover, leads to the loss of superconductivity as observed in $Mg_{1-x}Al_xB_2$ materials [3]. In our recent investigations of $Mg_{1-x}Al_xB_2$ ($0 \leq x \leq 1$) materials, measurements of physical properties and microstructure have revealed a rich variety of phenomena resulting from Al ordering with which we will be concerned in the following context.

$Mg_{1-x}Al_xB_2$ samples have been prepared by the conventional method. A detailed report on the sample preparation and x-ray analyses has been published in ref.8. Specimens for transmission-electron microscopy (TEM) observations were polished mechanically with a Gatan polisher to a thickness of around 50μm and then ion-milled by a Gatan-691 PIPS ion miller for 3 h. The TEM investigations were performed on a H-9000NA electron microscope with an atomic resolution of about 0.19nm.

Fig.1 (a) shows a series of the x-ray diffraction spectra of the $Mg_{1-x}Al_xB_2$ samples with x=0, 0.17, 0.5, 0.75, and 1.0, illustrating remarkable modifications of the 002 peaks. Two broad peaks for samples ranging from 0.09 to 0.25 have been observed as reported previously [3]. Our systematic study suggests that $Mg_{1-x}Al_xB_2$ materials with x ranging from either 0.09 to 0.25 or 0.70 to 0.8 exhibit complex structural and physical behaviors arising from phase separations. Direct evidences of the presence of two phases in these specific regions have been obtained by TEM investigations. Fig. 1(b) shows an electron diffraction pattern for x=0.17, indicating the coexistence of two structural phases in this material. The remarkable structural feature revealed in this pattern is the appearance of a new superstructure phase with noticeably different cell parameters as indicated. In order to characterize the essential origins of the structural anomalies in $Mg_{1-x}Al_xB_2$ system, we have performed an extensive structural study by means of x-rays diffraction, TEM observations and energy dispersive x-ray microanalysis. Aluminum ordering has been found to be a common phenomenon in the materials with x in the large range of 0.1 to 0.75. Especially, a new superstructure phase, with the optimal composition of $Mg_{0.5}Al_{0.5}B_2$ ($MgAlB_4$ phase), is discovered.

Fig.2 (a) shows an electron diffraction pattern taken from an $Mg_{0.5}Al_{0.5}B_2$ crystal, exhibiting the superstructure spots at the systematic (h, k, l+1/2) positions. Detailed analyses suggest that this superstructure originates from an ordered arrangement of Al and Mg layers along the c-axis direction. A brief schematic model for this superstructure is displayed in Fig. 2(b) clearly illustrating the atomic layers. Fig.2 (c) shows a high-resolution electron micrograph of $MgAlB_4$ crystal taken along the <010> zone-axis direction, exhibiting the atomic layers and superstructure along the c direction. This image was obtained from a thin region in the $MgAlB_4$ crystal; therefore, in combination with the results of theoretical simulations, the atomic structure in this superstructure phase could be identified. Image calculations, based on the schematic model of Fig. 2(b) together with the resultant structural distortions, were carried out by varying the crystal thickness and the defocus value. A calculated image with the defocus value of –0.45nm and the thickness of ~1.9nm is superimposed on the image, and appears to be in good agreement with the experimental one. Fig 2(d) shows the temperature dependent magnetization (upper panel) and resistivity (lower panel) for an $Mg_{0.5}Al_{0.5}B_2$ sample. The superconducting transition can be clearly recognized in either magnetization or resistivity with $T_c$ (onset) at around 12K.

Fig. 3(a) shows a brief diagram illustrating the $T_c$ as a function of the Al concentration (x). In this diagram, we have shown the results obtained from two sets of samples prepared under different of conditions; one was heated to the temperature of $850^oC$ for 2 hours and the other to $950^oC$ for 3hours. In order to facilitate our discussion, the average $T_c(x)$ are approximately displayed in this figure. The abrupt decrease of $T_c$ between x=0.25 and 0.4 is apparently in connection with the phase transformation from the conventional hexagonal structure to the superstructure [3]. In the range of x=0.40 to 0.6, the superconducting critical temperature, $T_c$, decrease gradually with the increase of Al concentration, and giving rise to a linear relationship between $T_c$ and x. Furthermore, in the vicinal region of x=0.5, $T_c$ decreases so slowly that we can obtain a clear plateau in $T_c$. This is quite analogous to the interesting behavior observed in the well-known high-$T_c$ superconductor $YBa_2Cu_3O_y$, in which oxygen stoichiometery, together with oxygen/vacancy ordering, significantly affects the physical properties and yields a plateau in $T_c$ at ~60K [9]. In-situ TEM observations show that the superstructure in the $Mg_{0.5}Al_{0.5}B_2$ sample is very stable without distinguishable changes in the temperature range of 100K up to 500K. It is also noted the superconducting transition depends markedly on the process of sample preparation, the superconducting transition can be either sharp or broad dependent on the synthesis conditions. The other evident decrease of $T_c$ occurs in the range of x=0.7 to 0.8, which is found to be in association with the disappearance of the superstructure as revealed in our TEM investigations. In other words, a phase transformation from the superstructure to $AlB_2$-like hexagonal structure occurs in this region. This transition seemingly appears in a narrow range neighboring x=0.75. Actually, we only observed clear evidence of phase separation in the x=0.75 material. A further study of the structural properties related to this phase transition is in progress. Materials with high Al concentrations of x>0.8 entirely lose superconductivity as measured in our experiments. Fig.3 (b) shows the composition dependence of the basic lattice parameters, a and c, for $Mg_{1-x}Al_xB_2$, evident anomalies become visible in the multiphase ranges of 0.09<x<0.25 and 0.7<x<0.8. Those coincide perfectly with the modifications of superconductivity. These facts suggest that the Al concentration and its ordered state significantly affect the physical properties of $Mg_{1-x}Al_xB_2$ materials and result in a rich variety of structural and physical issues. The further study of these issues may prove fruitful.

In conclusion, the $Mg_{1-x}Al_xB_2$ exhibits a variety of remarkable structural and physical properties in connection with Al ordering along the c-axis direction, such as structural transformations, phase separations, and the anomalous modifications in superconductivity. Though recent theoretical literatures about the electronic structure have performed for $Mg_{1-x}Al_xB_2$ and other related systems [10], the presence of the Al ordering in a large intermediate range of 0.1<x<0.75 will definitely lead to a further examination of the theoretical results.

Acknowledgments

The authors would like to express many thanks to Prof. G.C. Che, Miss S. L. Jia, Prof. D. N. Zheng, and Prof. L. P. You for their assistances. The work reported here was supported by "Hundreds of Talents" program organized by the Chinese Academy of Sciences, P. R. China.

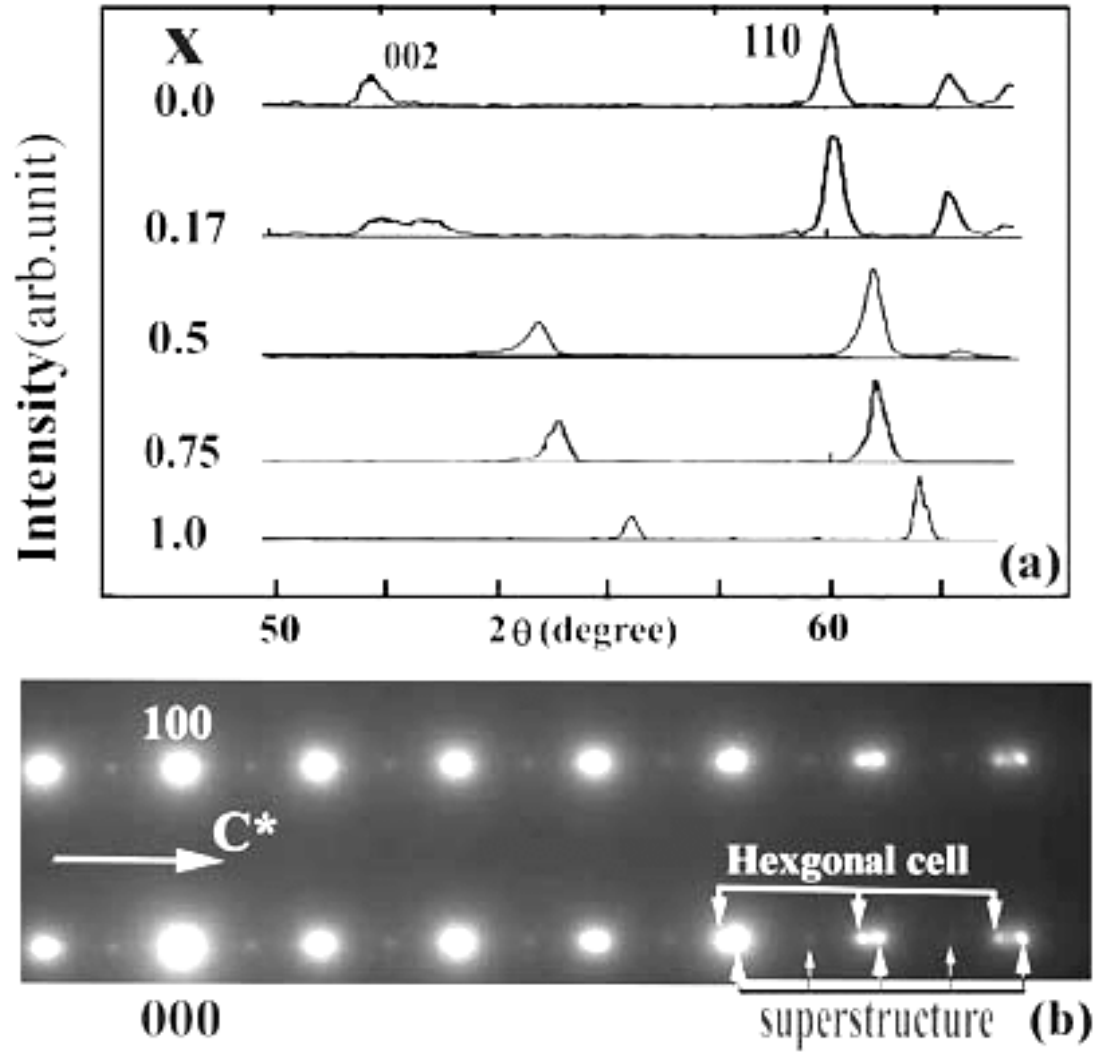

Fig. 1. (a) A series of x-ray diffraction spectra of $Mg_{1-x}Al_xB_2$, illustrating the evolution of (002)-reflection peak. (b) Electron diffraction pattern showing the presence of an additional superstructure phase in the x=0.17 sample.

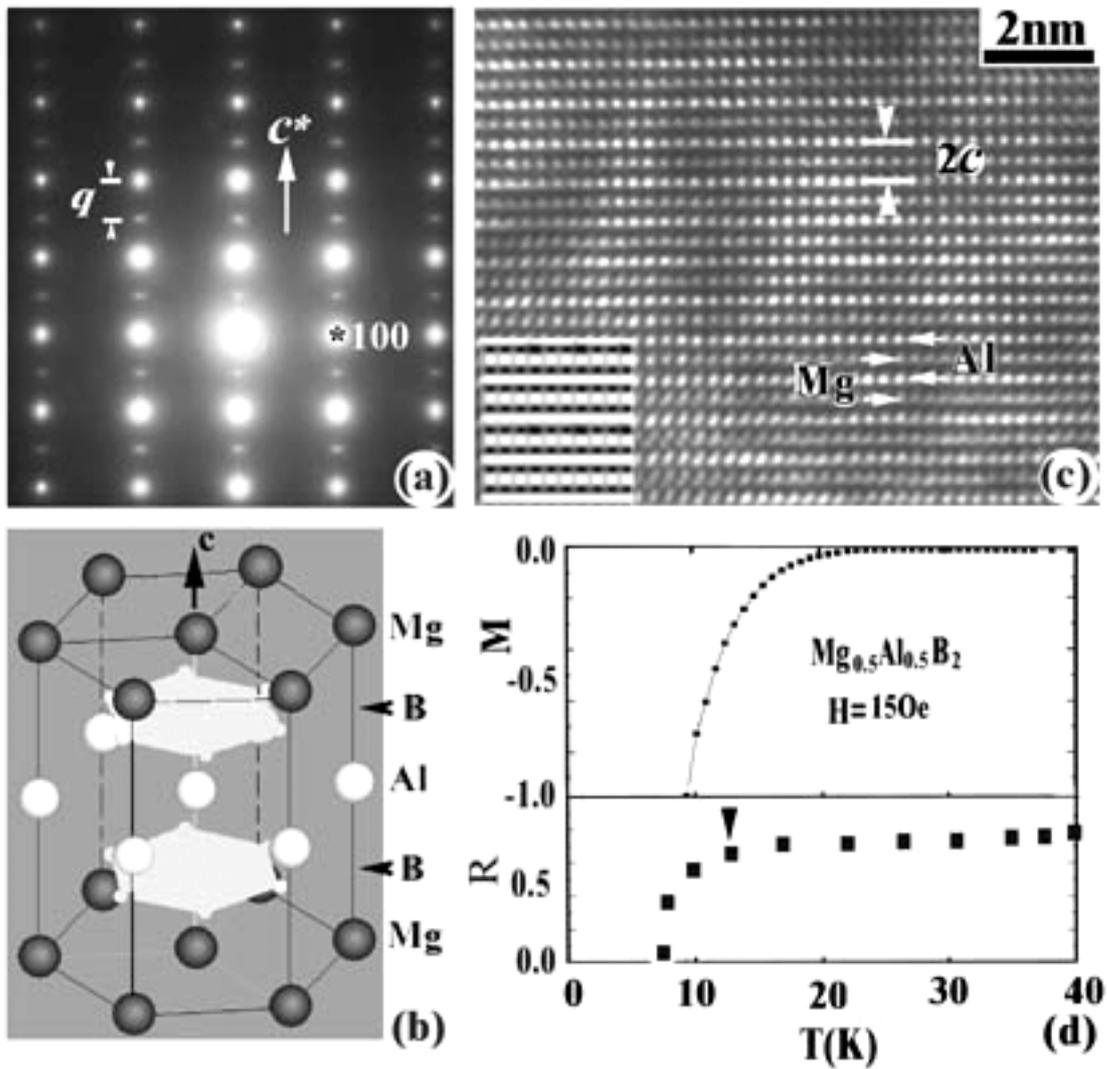

Fig. 2. (a) An electron diffraction pattern of the superstructure phase $MgAlB_4$, with the wave vector of $q=c^*/2$. (b) A schematic structural model of $MgAlB_4$. (c) High-resolution TEM image clearly exhibiting ordered Al and Mg layers along c direction. Inset shows a calculated image. (c) Temperature dependence of normalized magnetization (upper panel) and resistivity (lower panel) showing superconductivity in the $MgAlB_4$ sample.

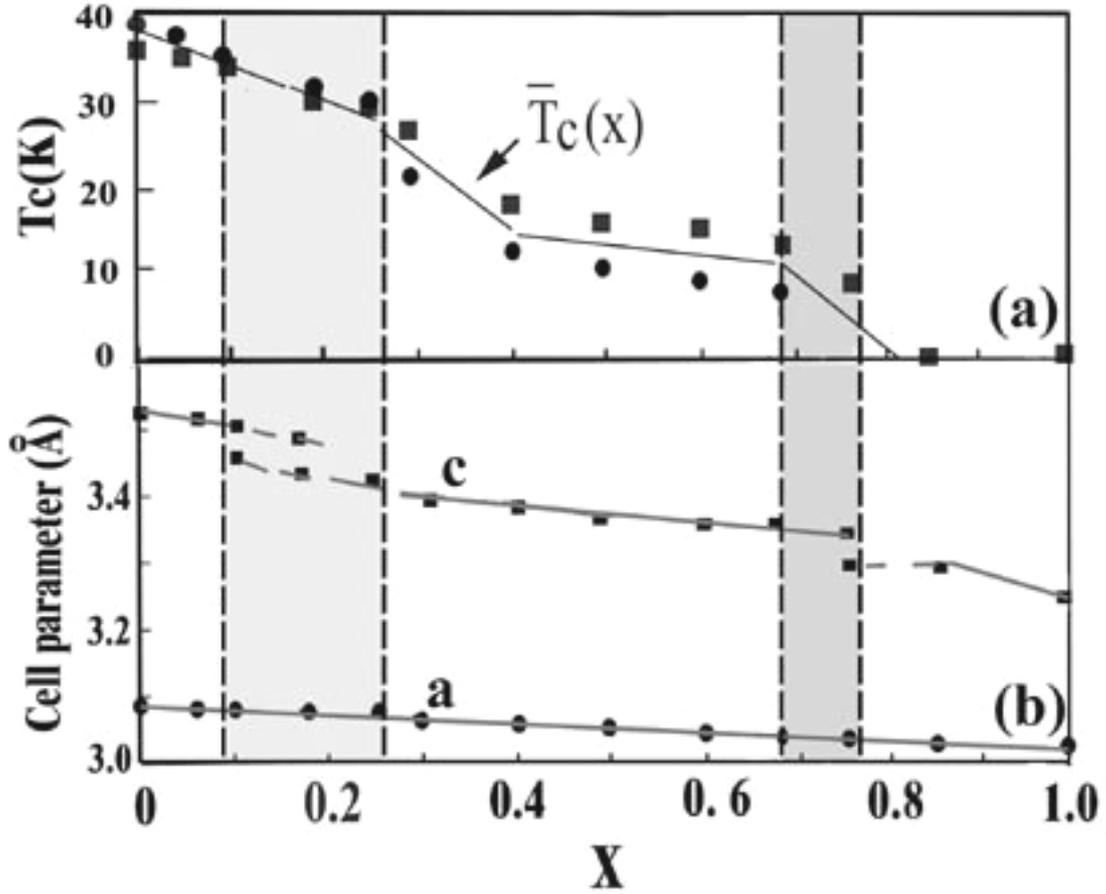

Fig. 3. Composition dependence of (a) superconducting temperature ($T_c$) and (b) basic structural parameters (*a* and *c*) for $Mg_{1-x}Al_xB_2$, anomalous structural and physical properties arising from Al ordering can be clearly recognized.